\documentclass[manuscript]{aastex}

\slugcomment{ }

\shorttitle{ Thermal Starless Core in the Orion A Cloud }
\shortauthors{Tatematsu et al.}

\begin{document}


\title{Thermal Starless Ammonia Core Surrounded by CCS\\
    in the Orion A Cloud}


\author{Ken'ichi Tatematsu,\altaffilmark{1,2}
Tomoya Hirota,\altaffilmark{1,2}
Satoshi Ohashi,\altaffilmark{3}
Minho Choi,\altaffilmark{4}
Jeong-Eun Lee,\altaffilmark{5}
Satoshi Yamamoto,\altaffilmark{6}
Tomofumi Umemoto,\altaffilmark{1,2}
Ryo Kandori,\altaffilmark{1}
Miju Kang,\altaffilmark{4}}

\and

\author{Norikazu Mizuno\altaffilmark{1,3}}


\altaffiltext{1}{National Astronomical Observatory of Japan, 
2-21-1 Osawa, Mitaka, Tokyo 181-8588, Japan; k.tatematsu@nao.ac.jp, 
tomoya.hirota@nao.ac.jp, umemoto.tomofumi@nao.ac.jp, 
r.kandori@nao.ac.jp, norikazu.mizuno@nao.ac.jp}
\altaffiltext{2}{Department of Astronomical Science, 
The Graduate University for Advanced Studies (SOKENDAI), 
2-21-1 Osawa, Mitaka, Tokyo 181-8588, Japan}
\altaffiltext{3}{Department of Astronomy, The University of Tokyo, Bunkyo-ku, Tokyo 113-0033, Japan; satoshi.ohashi@nao.ac.jp}
\altaffiltext{4}{Korea Astronomy and Space Science Institute, 
Daedeokdaero 776, Yuseong, Daejeon 305-348, South
Korea; minho@kasi.re.kr, mjkang@kasi.re.kr}
\altaffiltext{5}{School of Space Research, Kyung Hee University, 
Seocheon-Dong, Giheung-Gu, Yongin-Si, Gyeonggi-Do, 446-701, South Korea; jeongeun.lee@khu.ac.kr}
\altaffiltext{6}{Department of Physics, The University of Tokyo, Bunkyo-ku, Tokyo 113-0033, Japan; yamamoto@taurus.phys.s.u-tokyo.ac.jp}


\begin{abstract}
We imaged two starless molecular cloud cores, TUKH083 and TUKH122,
in the Orion A giant molecular cloud in the CCS and ammonia (NH$_3$) emission with the Very Large Array.  TUKH122 contains one NH$_3$ core ``TUKH122-n,'' which is elongated and has a smooth oval boundary.  Where observed, the CCS emission surrounds the NH$_3$ core. This configuration resembles that of the N$_2$H$^+$ and CCS distribution in the Taurus starless core L1544, a well-studied example of a dense prestellar core exhibiting infall motions. The linewidth of TUKH122-n is narrow (0.20 km s$^{-1}$) in the NH$_3$ emission line and therefore
dominated by thermal motions.  The smooth oval shape of the core boundary and narrow linewidth in NH$_3$ seem to imply that TUKH122-n is dynamically relaxed and quiescent.
TUKH122-n is similar to L1544 in the kinetic temperature (10 K), linear size (0.03 pc), and virial mass ($\sim$ 2 $M_{\odot}$). Our results strongly suggest that TUKH122-n is on the verge of star formation.  
TUKH122-n is embedded in the 0.2 pc massive (virial mass $\sim$ 30 $M_{\odot}$) turbulent parent core, while the L1544 NH$_3$ core 
is embedded in the 0.2 pc less-massive (virial mass $\sim$ 10 $M_{\odot}$) thermal parent core. 
TUKH083 shows complicated distribution in NH$_3$, but was not detected in CCS.  The CCS emission toward TUKH083 appears to be extended, and is resolved out in our interferometric observations.
\end{abstract}


\keywords{ISM: clouds
---ISM: individual (Orion Nebula, Orion Molecular Cloud)
---ISM: molecules
---ISM: structure---stars: formation}



\section{Introduction}
``Giant molecular clouds (GMCs)'' are known to form 
star clusters with both massive and low-mass stars while ``cold dark clouds''
(also called ``nearby dark clouds'', ``small molecular clouds'', or ``SMCs''; excluding ``infrared dark clouds'' here) can form only 
low-mass stars in isolation \citep[e.g.][]{shu87,tur88,ber07}.
These two categories of molecular clouds have different physical properties.
Molecular clouds have hierarchical structure, and structures having sizes of the order 0.1 pc are often called ``molecular cloud cores'' or simply ``cores.''
The nonthermal motion or turbulence is dominant in cores in GMCs,
while the thermal motion is more dominant in cores in cold dark clouds \citep[e.g.][]{tur88,tat93,cas95}.
Because most stars in the Galaxy form in GMCs, 
star formation from their cores is of great interest.  
How do stars form from turbulent cores in GMCs?
Do turbulent cores dissipate turbulence partially or completely 
before star formation?
Therefore, it is essential to observationally characterize cores in GMCs in
the turbulent environments where most stars are born.
It was theoretically suggested that cores are dynamical rather than quasistatic objects with relatively short lifetimes \citep[e.g.][]{bal03,bal07}.  
If this is true, cores in GMCs cannot be quasistatic, and will be far different from
quiescent cores in cold dark clouds \citep[e.g.][]{mye83b,pin11}.
Some researchers have pointed out 
that the dissipation of turbulence can initiate the star formation process \citep{mye83b,nak98,aso00}.  
However, it is not well established observationally how turbulent molecular clouds form stars.  We wonder whether stars form in dynamical cores or quiescent cores, whether turbulence dissipates before star formation, and how it does.  To know how star formation occurs in GMCs, it is essential to investigate starless molecular cloud cores that might lead to star formation in near future in detail.  

We take the Orion A cloud, which is the nearest and best studied GMC \citep[e.g.][]{gen89}. The overall structure is filamentary, and contains the ``$\int$-shaped filament'' (0.5 pc wide, 13 pc long) in the northern region \citep{bal87}.  \citet{joh99} mapped the $\int$-shaped filament at 450 and 850$\mu$m,
and found a chain of compact sources along a narrow ($<$ 0.14 pc) high-density filament.   The Orion A GMC has a hierarchical structure consisting of 1.4 pc-width filaments \citep{nag98} as well as 1 pc-sized clumps and 0.1 pc-sized cores \citep{bat83,bal87,cas90,tat93,ces94, chi97,tat98,lis98,aso00,joh06,ike07,tat08,tak13}. 

To identify candidate starless cores close to star formation, we use chemical evolution tracers.
\citet{tat93} mapped the Orion A GMC in CS $J$ = 1$-$0 with the Nobeyama 45 m radio telescope, and cataloged 125 cores
(we use the prefix of TUKH).
\citet{tat10} made single-pointing observations toward more than 60 TUKH cores in CCS and other molecules with the Nobeyama 45 m radio telescope.  They detected CCS in molecular cloud cores in the Orion A GMC for the first time; this molecule is known as a tracer of young molecular gas in cold dark clouds through observations and chemical model calculations \citep{suz92}.
TUKH122 and TUKH083 are the most and second-most intense cores in CCS $J_N$ = 4$_3-3_2$, respectively, in 
their observations.
From their data, TUKH083 is thought to be a young starless core, because it meets the category of the Carbon-Chain-Producing Regions (CCPRs) on the $N$(CCS)-$N$(NH$_3$)/$N$(CCS) diagram defined by \cite{hir09}.  On the other hand, TUKH122 is thought to be an older starless core, because it is not categorized in the CCPRs category.  \citet{tat14} mapped six TUKH cores in the Orion A GMC in N$_2$H$^+$ $J$ = 1$-$0 and CCS $J_N$ = 7$_6-6_5$ also with the Nobeyama 45 m telescope, and identified sub-peaks in these lines.  They confirmed that the column density ratio of N$_2$H$^+$ to CCS can be used as a chemical evolution tracer even in the GMCs if the gas temperature in the core does not exceed $\sim$ 25 K.  
The N$_2$H$^+$ sub-peaks (TUKH122C and TUKH122D) inside the TUKH122 region have the largest $N$(N$_2$H$^+$)/$N$(CCS) column density ratios (2$-$3) among the starless sub-peaks observed by \citet{tat14}.
The ratios are close to the criterion between starless and star-forming core peaks, which means these sub-peaks are close to star formation. TUKH122 
has the narrowest N$_2$H$^+$ line profiles in 
their N$_2$H$^+$ core samples.
In this study, we mapped two starless cores TUKH083 and TUKH122 in the Orion A 
GMC, by observing the ammonia (NH$_3$) and CCS emission to investigate the physical and chemical properties.
According to \citet{wil99}, TUKH083 and TUKH122 have 
NH$_3$ rotational temperatures of 35 and $<$ 15 K, 
respectively. These correspond to the kinetic temperature 
$T_K$ = 60 K and $<$ 15 K, respectively by using the conversion given by \citet{dan88}.
\citet{wil99} obtained $T_K$ also from CO $J$ = 3$-$2 peak intensities to be
15 and 12 K for TUKH083 and TUKH122, respectively. 
When we investigate the Herschel Observatory images, the dust continuum emission corresponding to TUKH122 is seen in Figure 2 of \citet{roy13}, but that corresponding to TUKH083 is not clear in Figure 10 of \citet{lom14} probably due to confusion in a crowded region.
According to Figure 2 of \citet{roy13}, the dust temperature toward TUKH122 is approximately 12$-$13 K.

In this work we adopt a distance of 418$\pm$6 pc based on
the work of \cite{kim08} as the best estimate; at this distance 1' corresponds to 0.12 pc 

\section{Observations}


Observations were carried out by using the Karl G. Jansky Very Large Array (hereafter VLA)
of the National Radio Astronomy Observatory\footnote{The National Radio Astronomy Observatory is a facility of the National Science Foundation operated under cooperative agreement by Associated Universities, Inc.} of the United States of America in the D configuration.
The observation dates are five days between 2010 August 29 and 2010 September 9.  
We observed NH$_3$ $(J, K)$ = (1, 1) at 23.694495 GHz \citep{ho83}
and 
CCS $J_N$ = 4$_3-3_2$ at 45.379033 GHz \citep{yam90}
in both of the right-hand and left-hand circular polarizations,
simultaneously.
We employed the receiver front-ends for
1.3 cm K band and 0.7 cm Q band.
The tracking velocity was $v_{LSR-K}$ = 4.0 km s$^{-1}$.
The receiver back-end was the WIDAR correlator in the OSRO2 mode.
The channel width 
was 15.625 kHz (corresponding to 0.197 km s$^{-1}$) for NH$_3$ 
and
31.25 kHz  (corresponding to 0.207 km s$^{-1}$) for CCS.
The number of the frequency channels was 256. 
We observed 3C147 as the flux and bandpass calibrator.  
The observed intensity is reported in terms of Jy beam$^{-1}$.
We used J0541-0541 as the phase calibrator.
Immediately before the flux, bandpass, and phase calibration,
we made pointing observations in X band toward the same calibrators.
Tables 1 and 2 list observation sessions and observational parameters.
The observed data were flagged, calibrated, and imaged (CLEANed) using
the CASA 4.1 software package. 
The flux of the flux calibrator 3C147 is assumed
to be 1.783 Jy at K band and 1.0136 Jy at Q band, 
from the model
of ``Perley-Butler 2010'' in CASA.
The CLEAN deconvolution was made using the simple Clark
Algorithm with the natural weighting and with UV taper.
For Q-band observations, we applied the Gaussian filter 
of ``outertaper'' = 3$\farcs$0 in addition
in the CLEAN deconvolution.
The primary beam size was about 2$\arcmin$ for K band and 1$\arcmin$ for Q band.
We used the software package AIPS to plot images and spectra 
to be presented in this paper.
Figures 1 and 2 show the primary beams of our interferometric observations at K band toward TUKH083 and TUKH122 superimposed on 
the CS $J$ = 1$-$0 maps of \citet{tat93}.
All the interferometric maps are corrected for the primary beam response.




\section{Results}
Core TUKH122 was clearly detected in both the NH$_3$ and CCS emission.  
Core TUKH083 was detected only in the NH$_3$ emission.
We have not detected the continuum emission either in K or Q bands.  The 3$\sigma$ upper limit to the continuum emission 
is 1.56 mJy beam$^{-1}$ or 0.276 K for K band and  5.4 mJy beam$^{-1}$ or 0.189 K for Q band.

\subsection{TUKH122 map}
Figures 3 to 5 show NH$_3$ and CCS velocity-integrated intensity maps toward TUKH122. 
Figure 4 shows the NH$_3$ map.
The NH$_3$ map shows the emission distribution elongated in the northwest-southeast direction.  
The boundary shape is very smooth and oval.
The NH$_3$ emission roughly corresponds to the N$_2$H$^+$ emission observed by \cite{tat14} as TUKH122C and D.
The noise level increases outward, because the primary beam response is corrected.  All the features outside the primary beam size are noise rather than the emission.
The NH$_3$ core has some internal structure, and the southeastern side of the NH$_3$ core looks clumpy.  It might be affected by higher noise level near the edge of the primary beam to some extent, but it is likely that some internal structures are real.  
We call the overall core TUKH122-n, and internal sub-cores TUKH122-n1 and n2.
The direction of the elongation of TUKH122-n is more or less in parallel with
the CS filament (P.A. $\sim$ 135$\arcdeg$) 
having a width of $\sim$ 0.1 pc in Figure 2.
Table 3 lists the physical parameters of TUKH122-n, n1, and n2.  We fit two-dimensional Gaussians to the emission distribution, and derive the beam-deconvolved parameters.  $a$, $b$, and PA  are the major and minor diameters, and the position angle, respectively.
The axial ratio (minor diameter/major diameter) is 0.26, which is smaller than the typical value in cores in cold dark clouds, on the order 0.5 \citep{mye91}. 
This core elongation is 
also in parallel with the global filamentary shape \citep[e.g.][]{bal87,tat93} of the Orion A GMC.
The flux of the core TUKH122-n obtained through the spatial 
two-dimensional Gaussian fitting is 392 mJy km s$^{-1}$.
Using the MPIfR 100m radio telescope,
\citet{wil99} observed TUKH122 in NH$_3$, and obtained 
a main-beam radiation temperature of $T_R$ = 2.28 K 
and an FWHM linewidth of 0.8 km s$^{-1}$, resulting in 
an integrated intensity of 1.94 K km s$^{-1}$.  
This corresponds to 1.64 Jy km s$^{-1}$
in their 43$\arcsec$ beam.
Then, 24\% of the single dish flux was resolved with the VLA.

Our interferometric CCS $J_N$ = 4$_3-3_2$ observations show three sub-cores, TUKH122-c1, c2, and c3 (Figure 5). 
Table 3 lists them. These emission peaks coincide well with the single-dish CCS $J_N$ = 7$_6-6_5$ peaks in \citet{tat14}: TUKH122-c1 roughly corresponds to their TUKH122B, and TUKH-c2 and -c3 correspond to their TUKH122.
The sum of the TUKH122-c1, c2, and c3 flux is 209 mJy km s$^{-1}$.
The single-dish observations toward TUKH122 \citep{tat10} show that the CCS $J_N$ = 4$_3-3_2$ intensity and linewidth are $T_A^*$ = 0.47 K or $T_R$ = 0.78 K and an FWHM linewidth of 0.55 km s$^{-1}$, respectively, resulting in 
an integrated intensity of 0.4 K km s$^{-1}$.  
This corresponds to 1.0 Jy km s$^{-1}$ in their 38$\farcs$5 beam.
Our interferometric observations resolved 21\% of the emission.
The CCS emission was detected in two velocity bins (2 $\times$ 0.2 km s$^{-1}$) in the VLA observations.  It is hard to measure the linewidth, but it will be $\sim$ 0.4 km s$^{-1}$ or less.
The single-dish CCS observations \citep{tat14} also show the emission at R.A. (J2000.0) = 5$^h$39$^m$46$\fs$5, Dec. (J2000.0) = $-$7$\arcdeg$30$\arcmin$49$\arcsec$ (TUKH122E), although contours are not closed due to the limited observation area.
This peak is located outside the CCS observation field of the present study, but  just 30$\arcsec$ southeast from the NH$_3$ core boundary. Then the NH$_3$ core has the VLA CCS sub-cores on one side and the single-dish CCS on the other side, along its elongation.  It is most likely that the CCS emission surrounds the NH$_3$ core.
This is very similar to the cases of cores in cold dark clouds,
the N$_2$H$^+$-CCS configuration in L1544 and NH$_3$-CCS configuration in L1498 \citep{aik01,lai00}.
L1544 is a dense prestellar core exhibiting infall motions \citep{taf98}. 
\cite{aik01} calculated numerical chemical models for a collapsing core, and found that models are consistent with the observations of L1544: the collapsing core has a chemically evolved N$_2$H$^+$ inner core surrounded by the chemically young CCS region. This likely also explains the NH$_3$ and CCS configuration in TUKH122.
TUKH122-n, L1544, and L1498 have similar linear sizes in pc.  We wonder about the current status of TUKH122-n regarding the core stability. From the similarity between TUKH122-n and L1544, it is possible that TUKH122-n may be unstable now, or may become unstable in near future by dissipating the remaining nonthermal motion or by accreting surrounding gas after turbulence dissipation. 
Taking also into account high $N$(N$_2$H$^+$)/$N$(CCS) ratio implying an evolved starless core \citep{tat14}, it is most likely that TUKH122-n is a thermal starless core on the verge of star formation.  

Our velocity resolution of 0.197 km s$^{-1}$ is insufficient to allow us to see whether TUKH122 is collapsing \citep{zho93,zho95,cho95,taf98} or has other systematic motion such as oscillation \citep{lad03,ket14,chi14}.
In the single-dish N$_2$H$^+$ spectra of \cite{tat14} at a resolution of  0.1 km s$^{-1}$  toward TUKH122, there is no hint of the blue-skewed spectrum of protostellar collapse.  However, it is highly desirable to obtain better velocity resolution data to investigate whether TUKH122 is collapsing or has other systematic motions.

\subsection{TUKH083 map}
Figure 6 shows the NH$_3$ map toward TUKH083. 
The emission is very clumpy.
Table 3 lists 5 NH$_3$ sub-cores TUKH083-n1 to n5.
A feature at R.A. (J2000.0) = 5$^h$36$^m$50$\fs$5, Dec. (J2000.0) = $-$6$\arcdeg$31$\arcmin$47$\arcsec$ is located near the edge of the primary beam, and is as compact as the synthesized beam. Because its intensity is close to 3$\sigma$, we do not catalog it as reliable detection.
The overall complicated shape may suggest that the core is not dynamically 
relaxed yet. 
We did not detect the CCS emission toward TUKH083, although this source is the second most intense CCS $J_N$ = 4$_3-3_2$ core in \cite{tat10}.
The 3$\sigma$ upper limit at 0.4 km s$^{-1}$ resolution is 3.6 mJy beam$^{-1}$ km s$^{-1}$.
This means that the CCS $J_N$ = 4$_3-3_2$ emission toward TUKH083 is extended and resolved out in our interferometric observations.

\subsection{NH$_3$ spectra}
Figures 7 and 8 show the NH$_3$ spectra toward TUK122 and TUKH083,
respectively.   
See \citet{mye83} for details of the hyperfine components of 
NH$_3$ (1, 1) and their feature names (features 1 to 8).
NH$_3$ hyperfine feature 1 is located outside our effective spectral window.
NH$_3$ hyperfine feature 4 is the main component and the other 
features are satellite components.
The intrinsic relative intensity of hyperfine components 
is 0.52, 0.38, 0.56, 0.33, 0.48, and 0.28 for satellite features 
2, 3, 5, 6, 7, and 8
with respect to the main component feature 4, respectively \citep{rydbeck77,ho83}.
Figure 7 shows the hyperfine component intensities toward TUKH122.
Hyperfine fitting tells us that the optical depth of the main component
is derived to be as large as 2.
This is in contrast with
the NH$_3$ (1, 1) spectrum obtained 
toward the dark cloud core L1498 by \citet{mye83}, which looks optically thinner.
TUKH083 (Figure 8) shows only NH$_3$ hyperfine features 4 and 5, and their 
observed intensity ratio is close
to the intrinsic relative intensity ratio.
Then, even the main component is optically thin.

\subsection{Velocity structure in NH$_3$}
Figures 9 and 10 show the intensity-weighted radial velocity maps toward TUKH122
and TUKH083, respectively. The radial velocity in TUKH122 is fairly constant over the core.
We investigate the velocity field toward TUKH122-n in NH$_3$ 
in detail.
To see the velocity structure along the elongation,
we rotate the image data anticlockwise by 33$\fdg$3 so that
the elongation becomes vertical.
Then, we average the data along the x axis 
(perpendicular to the elongation), and 
obtain the position-velocity diagram.
Figure 11 illustrates the position-velocity diagram of 
NH$_3$ hyperfine features 7 and 8 toward TUKH122.
If all the hyperfine components are optically thin, 
we expect to see spatial components
aligned parallel (horizontally) with each other.
These hyperfine components do not show such correspondence well.
Then, the optical depth probably affects the intensity distribution
in these position-velocity diagrams.
We derive the linewidth by using optically thinner line,
NH$_3$ hyperfine feature 8.
By fitting a two-dimensional Gaussian to the NH$_3$ hyperfine feature 8 in Figure 11,
we obtain an FWHM linewidth
of 0.280$\pm$0.002 km s$^{-1}$.
Deconvolved with an instrumental velocity resolution of 0.197 km s$^{-1}$,
the intrinsic FWHM linewidth is 0.20 km s$^{-1}$.
This value is close to the 
thermal FWHM linewidth $\Delta v$(th) = (8(ln 2)$kT_k/m)^{1/2}$ for NH$_3$ at $T_k$ = 10 K of 0.16 km s$^{-1}$. 
Here, $k$ is the Boltzmann constant, and $m$ is the mass of the molecule.
The FWHM linewidth for the mean mass per particle (2.33 a.m.u.) at 10 K is 0.44 km s$^{-1}$.
The turbulent FWHM linewidth $\Delta v$(turb) and total FWHM linewidth $\Delta v$(tot) \citep{mye83} are derived to be 0.16 and 0.47 km s$^{-1}$, respectively.
Because $\Delta v$(th) $>$ $\Delta v$(turb), TUKH122-n is thermally dominated.
If there is a velocity gradient across (perpendicular to) the elongation, it may contribute to $\Delta v$(turb), but we do not see any hint of such a gradient across the elongation.
We observe a slight velocity gradient
of 3.6 km s$^{-1}$ pc$^{-1}$ along the elongation from feature 8 in Figure 11.
Then, the specific angular momentum $J/M = (2/5) R v_{rot}$ (2/5 for a sphere) is 
1.2 $\times$ 10$^{-3}$ km s$^{-1}$ pc. 
This follows the empirical best-fit $J/M-R$ relationship obtained by
\citet{goo93} for $R$ = 0.03$-$0.3 pc, or is located slightly lower (smaller $J/M$).
We derive the ratio $\beta$ of rotational energy to the gravitational energy.  We adopt the definition of $\beta$ 
= (1/3) $\omega^2 R^3/GM$ \citep{goo93}, and obtain $\beta$ = 1.6 $\times$ 10$^{-2}$.
Here, $\omega$ is the angular speed, and $G$ is the gravitational constant.
This small value ($<<$ unity) is close to the typical value of 0.02 obtained in dark cloud cores \citep{goo93}, and indicates that the core is not rotationally supported.
Because the axial ratio of the core TUKH122-n is as small as 0.26 and rotation cannot account for core elongation,
it is likely that magnetic fields play a key role in
the core shaping of TUKH122.
\cite{nak98} has concluded that molecular clouds are magnetically supercritical.
\cite{cru99} has shown that the mass-to-magnetic flux ratio is about twice as critical, and static magnetic fields and turbulence (MHD waves) are equally important in cloud energetics.  He has concluded that magnetic fields are important in the physics of molecular clouds.  Our observations seem to be consistent with his conclusion.

Next, we look into the case of TUKH083.
Figure 12 shows the position-velocity diagram along right ascension 
toward TUKH083 of NH$_3$ hyperfine feature 4.
The emission is integrated along declination within a box containing the
detected NH$_3$ emission.
Precisely, feature 4 consists of eight hyperfine components, and the two most intense components have a frequency difference corresponding to 0.32 km s$^{-1}$ \citep{rydbeck77}.
However, for the TUKH083 NH$_3$ core, 
it is hard to separate hyperfine components in feature 4 because of overcrowding.

\subsection{Comparison between TUKH122, TUKH 123, and the dark cloud core L1544}
We compare the physical parameters of our cores in Orion and the dark cloud core L1544 in Taurus.
To the east of TUKH122, a cluster-forming core TUKH123 accompanying 
L1641-south3 \citep{fuk89} containing six protostars \citep{meg12} is located (Figure 2).
Table 4 summarizes the physical parameters of TUKH122 and TUKH123 in the Orion A GMC, and that of L1544 in the dark cloud.
The CS $J$ = 1$-$0 data of TUKH122, and TUKH123 are taken from
\cite{tat93},
and the data of L1544 is taken from
\cite{ben89,taf98}.
We adopt the NH$_3$ rotation temperature for Orion cores from
\citet{wil99}, and converted it to the kinetic temperature $T_k$ using
the relation given by \citet{dan88}.
For TUKH122 and TUKH123, \citet{wil99} list upper limits of $T_k <$ 15 K, and
we adopted $T_k$ = 10 K assuming that they have temperatures similar to
those in cores in cold dark clouds.
Regarding the radius $R$, we adopt the HWHM (= FWHM/2) radius.
The major and minor radius at the half-maximum level are deconvolved values for the synthesized beam.
Then, the radius $R$ is calculated from a geometrical mean of the major and minor radius.
Regarding linewidth, we adopt the definition of the total linewidth $\Delta v$ (tot) from \citet{mye83b},
which takes into account the difference between the mass of the observed molecule and the mean mass per particle.
We use the virial mass instead of the LTE mass here, because the molecular abundance is not precisely known.  
The virial mass is derived by using the total linewidth instead of the observed linewidth, and therefore is slightly different from those reported in \cite{tat93} and \cite{taf98}.
The gas density is derived from the radius and virial mass, and defined as $n$ = 1.2 $n$ (H$_2$) by taking into account the contribution of He.
We list the free-fall time estimated from density.
Table 4 shows that TUKH122-n in NH$_3$ and L1544 in N$_2$H$^+$ are very similar in the kinetic temperature (10 K), linear size (0.03 pc radius), and mass ($\sim$ 2 $M_{\odot}$).
It was often suggested that cores in GMCs and those in cold dark clouds are different in the physical properties \citep[e.g.][]{tur88}, but our observations revealed that the Orion A GMC contains at least one core very similar to the dark cloud core.

\section{Discussion}
\subsection{Turbulence in cores}
The NH$_3$ cores of TUKH083 and TUKH122 are embedded in the CS core observed by \citet{tat93}.
The thermal Jeans length and mass of the CS core are calculated 
for the adopted kinetic temperature $T_k$ (table 3).
The virial mass of the TUKH122-n core 1.5 $M_{\odot}$ is close to
the Jeans mass in the parent turbulent CS core.
This is consistent with an idea that TUKH122-n has been formed as gravitational instability in the parent turbulent equilibrium core.

The intrinsic NH$_3$ linewidth in our interferometric observations is as small as 0.20 km s$^{-1}$.
The single-dish CCS $J_N$ = 7$_6-6_5$ and N$_2$H$^+$ linewidth 
are 0.27$-$0.39 and 0.27$-$0.28 km s$^{-1}$, respectively \citep{tat14}, and are almost thermal.
On the other hand, the single-dish CS $J$ = 1$-$0 linewidth toward TUKH122 is as large as 0.83 km s$^{-1}$ \citep{tat93}.
According to \cite{tat98} the optical depth of this line is $<$ 0.9,
and therefore the linewidth will not be seriously affected.
Figure 13 shows the CS $J$ = 1$-$0 velocity channel map of \cite{tat93} toward TUKH122.
The CS emission is centered at the field center of the NH$_3$ and CCS observations, from $v_{LSR-K}$ = 3.6 to 4.2 km s$^{-1}$.  Then, it is observed as a non-thermal, turbulent core.
In our observations, the CCS emission is observed only within a range of 0.4 km s$^{-1}$ toward TUKH122,
which is close to the thermal FWHM linewidth.
These facts will imply turbulence dissipation in TUKH122: the turbulent TUKH122 CS core 
($R$ = 0.18 pc, $\Delta v$(tot) = 0.94 km s$^{-1}$)
has formed the quiescent, thermal NH$_3$ core associated with CCS inside.
The viral mass and corresponding column density of the TUKH122 CS core is 33 $M_{\odot}$ and $N$ (H$_2$) = 2.2 $\times$ 10$^{21}$ cm$^{-2}$, respectively.
This column density is consistent with that obtained from dust continuum observations with the Herschel Observatory \citep{roy13}, $N$ (H$_2$) = 1.6$-$1.9 $\times$ 10$^{21}$ cm$^{-2}$ from their figure.
Furthermore, the NH$_3$ core looks dynamically relaxed, reminding
us of quasistatic evolution.
The NH$_3$ and CS cores are located in the CS filament (Figure 2).
We suggest that the turbulent equilibrium CS cores are formed from the turbulent filamentary molecular cloud due to the gravitational instability \citep{han93}, and the dynamically relaxed thermal NH$_3$ cores are then formed slowly with turbulence dissipation and gravitational instability.

Although the TUKH122 and L1544 NH$_3$ cores are very similar in terms of the kinetic temperature, linear size, and mass, they differ in terms of the mass of parent core: TUKH122 has a turbulent massive ($\sim$ 30 $M_{\odot}$) parent core observed, while L1544 has a thermal less-massive ($\sim$ 10 $M_{\odot}$) parent core.  Indeed, it is known that 0.1 pc-sized cores in the Orion A GMC have systematically larger linewidths and larger masses than similar-size cores in cold dark clouds \citep{tat93}.  It is possible that such a difference is related to an ability for GMC cores to form star clusters. In dark cloud cores, \cite{goo98,lai00} investigated the single-tracer single-cloud linewidth-size relation, and concluded that linewidth becomes coherent (thermal-motion dominated constant values against radius) on length scales less than $\sim$ 0.1 pc.  \cite{pin10} concluded that a transition to coherence occurs at 0.04 pc at B5 in the Perseus cold dark cloud.  GMC cores have different linewidth-size relation compared with dark cloud cores \cite{tat93,cas95}. Judging from Figure 6 of \cite{cas95}, the transition in GMC cores may occur at a smaller radius (0.01$-$0.1 pc).   This is consistent with our observations.

\subsection{Comparison between TUKH083 and TUKH122}
We found that TUKH083 and TUKH122 are very different in the distribution of the 
NH$_3$ emission.
TUKH083 is clumpy, while TUKH122 has a smooth oval boundary.
In CCS, TUKH122 was detected in the interferometric observations, but TUKH083 was not detected.  This means that the CCS emission distribution is more compact in TUKH122 than in TUKH083.
There are two possibilities.
First, because TUKH083 is warmer than TUKH122,
different temperatures caused different core formation modes.
Second, TUKH083 is younger, and has not been dynamically relaxed yet.
From $N$(NH$_3$)/$N$(CCS) and $N$(N$_2$H$^+$)/$N$(CCS) \citep{tat10,tat14},
TUKH083 is most likely younger than TUKH122.
However, clearly we need more samples to establish the picture of core evolution in various environments in GMCs.

\section{Summary} 
We mapped two starless molecular cloud cores, TUKH083 and TUKH122,
in the Orion A GMC in CCS and NH$_3$ with the VLA.  The TUKH122 NH$_3$ core (TUKH122-n) is elongated, and has a smooth oval boundary.  Where observed, the CCS emission surrounds the NH$_3$ core.  This is very similar to N$_2$H$^+$-CCS distribution in the Taurus cores L1544 and L1498.  TUKH122-n has an NH$_3$ linewidth of 0.20 km s$^{-1}$, which implies that the core is thermally supported.  The linewidth and shape of the NH$_3$ core seem to represent a dynamically relaxed quiescent core.  TUKH122-n (associated with CCS) is embedded in the parent turbulent CS core, which suggests TUKH122-n has formed through turbulence dissipation.
The parent core of TUKH122-n is three times more massive than that of L1544, which may suggest difference in the capability of forming the cluster.
The physical parameters of TUKH122-n resemble those of the L1544 NH$_3$ core.  The virial mass of the TUKH122-n core is 1.5 $M_\odot$,
and is on the order of the Jeans mass of the parent CS core.  TUKH083 was detected in NH$_3$ showing complicated distribution, but not in CCS.    The CCS emission toward TUKH083 appears to be resolved out.



\acknowledgments
K.T. is grateful to Tomoyuki Hanawa for comments on the
draft.
The authors would like to thank an anonymous referee for helpful comments.
Data analysis were carried out on the common use data analysis computer system at the Astronomy Data Center, ADC, of the National Astronomical Observatory of Japan.
M.C. was supported by
the Core Research Program of National Research Foundation
funded by the Ministry of Science, ICT and Future Planning
of the Korean government (grant number NRF-2011-0015816).
J.-E.L. was supported by the Basic Science Research Program through the National Research Foundation of Korea (NRF) funded by the Ministry of Education of the Korean government (grant number NRF-2012R1A1A2044689).  



{\it Facility:} \facility{VLA}.

\clearpage



\begin{figure}
\epsscale{.80}
  \begin{center}
    \plotone{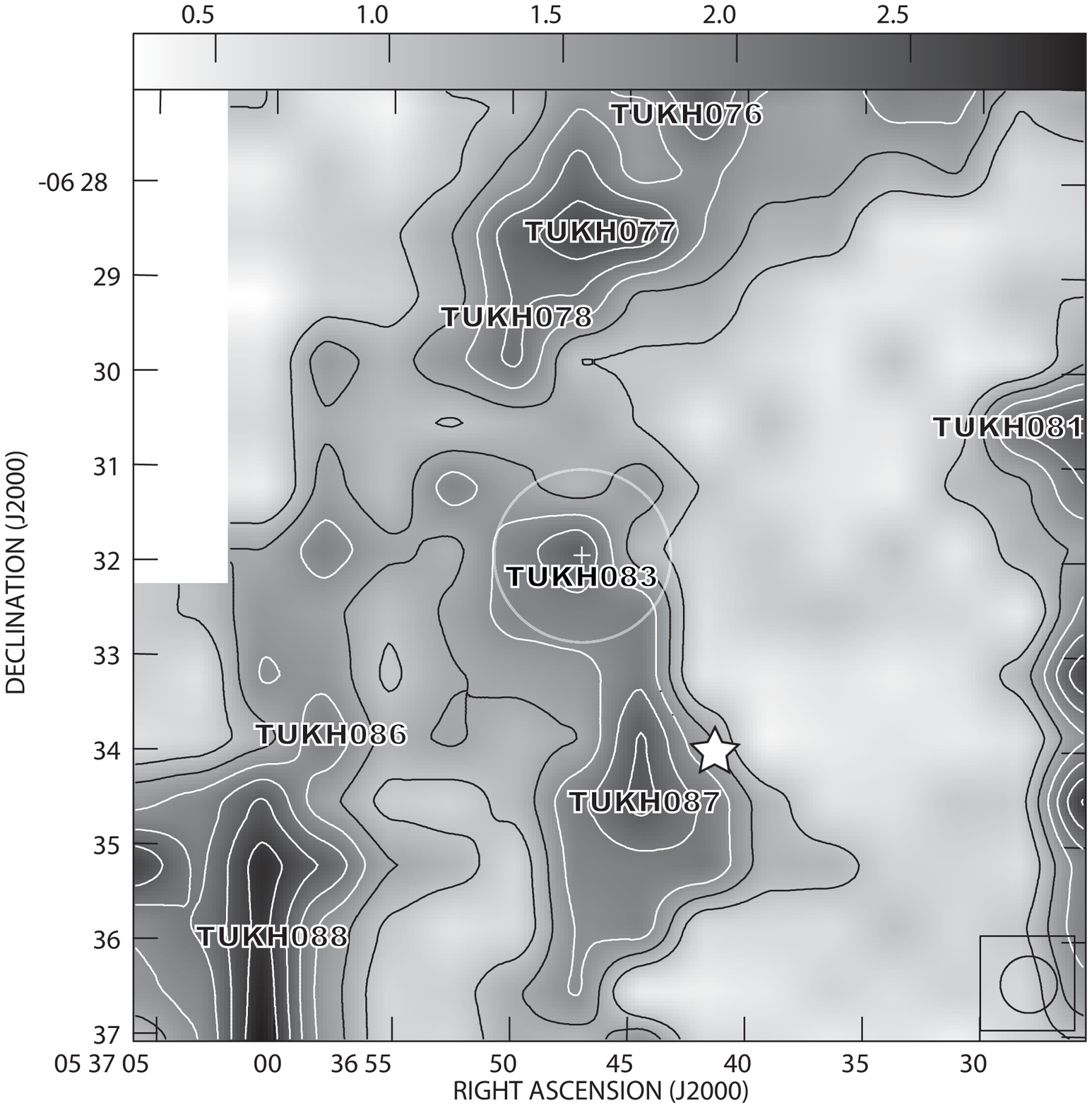}
  \end{center}
  \caption{
CS $J$ = 1$-$0 peak intensity map around TUKH083 taken from \citet{tat93}.
The contour starts at 3$\sigma$ with an interval of 1$\sigma$, where
1$\sigma$ is $T_A^*$ = 0.34 K or the main-beam radiation temperature $T_R$ = 0.57 K.
The position of the Spitzer protostar \citep{meg12} is shown with the star sign.
The primary beam (field of view) of the VLA observations at K band is shown as a white 
circle centered at the plus sign.
The primary beam at Q band is twice smaller.
The beam size for the CS observations is shown in the lower right corner.
}\label{fig:figure1}
\end{figure}

\clearpage

\begin{figure}
\epsscale{.80}
  \begin{center}
    \plotone{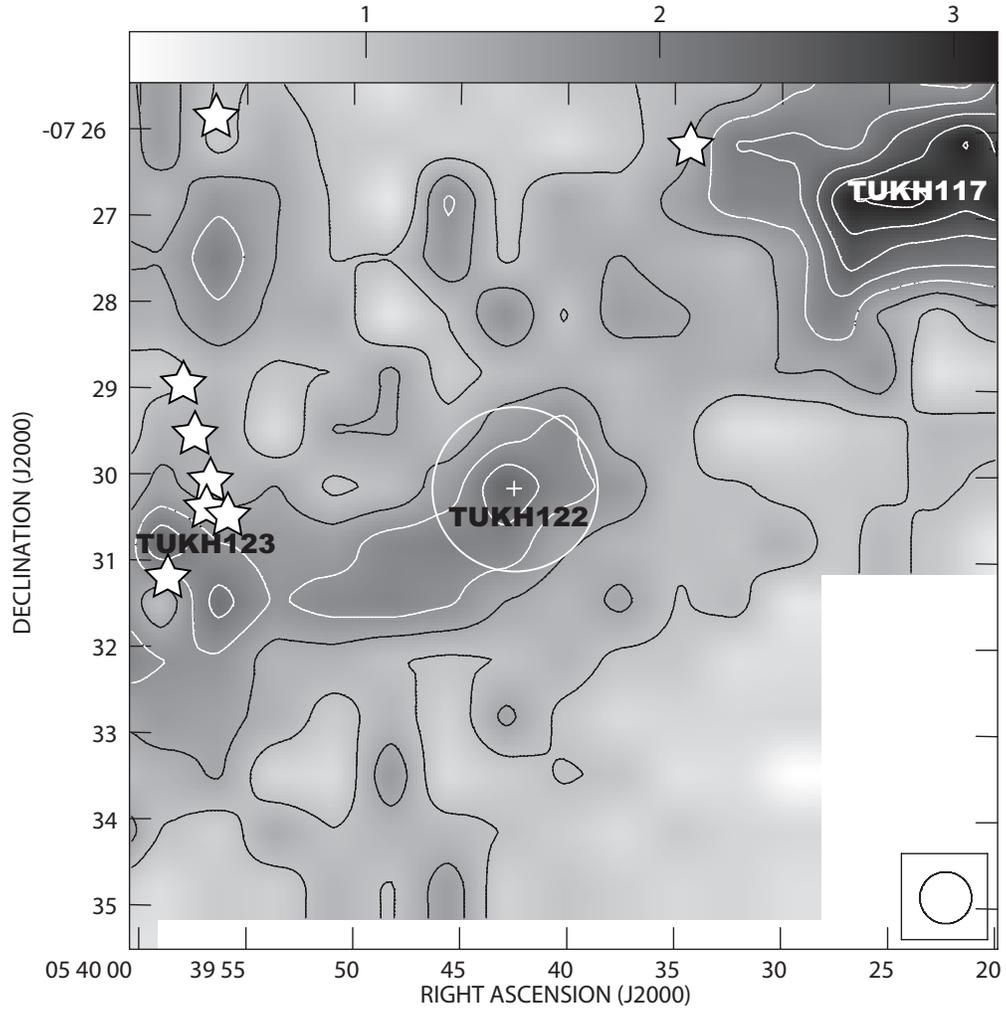}
  \end{center}
  \caption{
The same as Figure 1 but for TUKH122.
The cluster of Spitzer sources associated with core TUKH123 
is L1641-south3 \citep{fuk89}.
}\label{fig:figure2}
\end{figure}

\clearpage

\begin{figure}
\epsscale{.80}
  \begin{center}
    \plotone{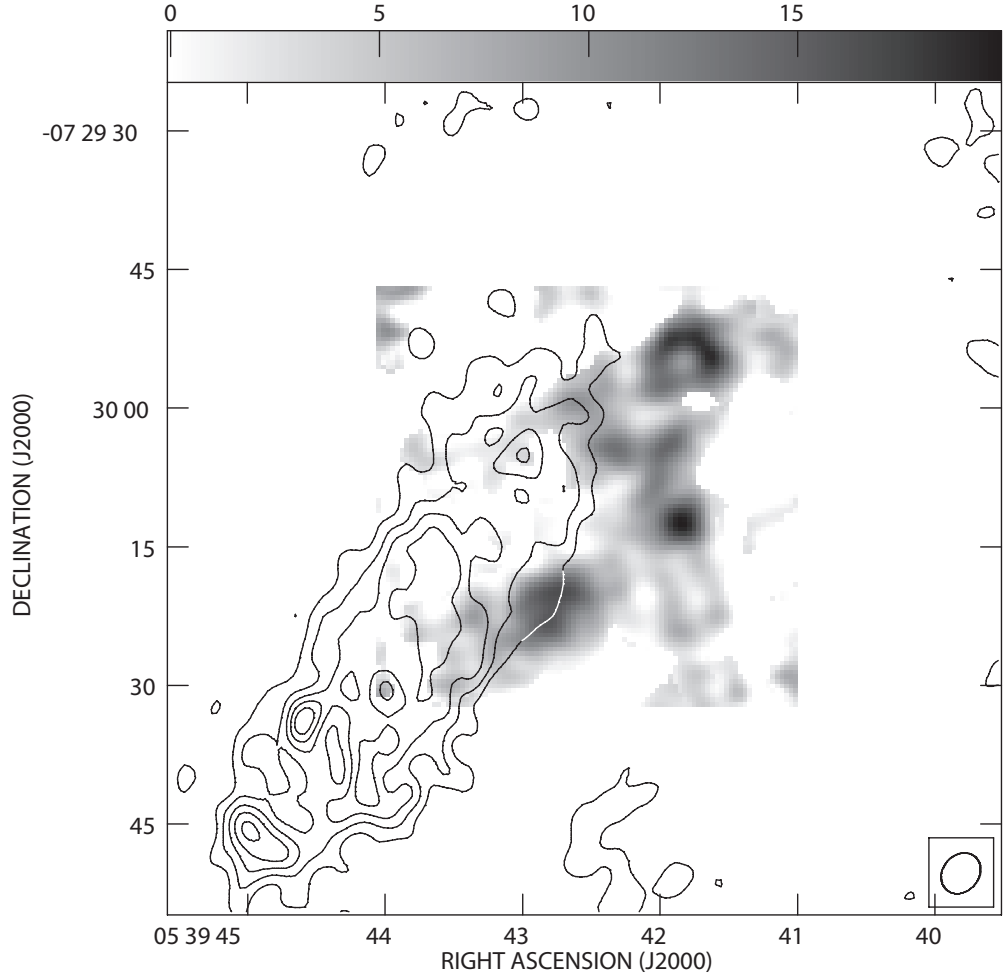}
  \end{center}
  \caption{
CCS integrated intensity map (gray scale) superimposed on NH$_3$ integrated intensity map (contours) toward TUKH122.  The lowest contour level and contour interval are 2.4 mJy beam$^{-1}$ km s$^{-1}$.
The maximum intensity for the gray scale and the contour are 19.8 and 14.3 mJy beam$^{-1}$ km s$^{-1}$, respectively.  In the lower right corner, the synthesized beam for CCS is shown.
That for NH$_3$ has a similar size.
The CCS emission was observed only in the central area (see Figures 4 and 5 for the detailed observed areas).
}\label{fig:figure3}
\end{figure}

\clearpage

\begin{figure}
\epsscale{.80}
  \begin{center}
    \plotone{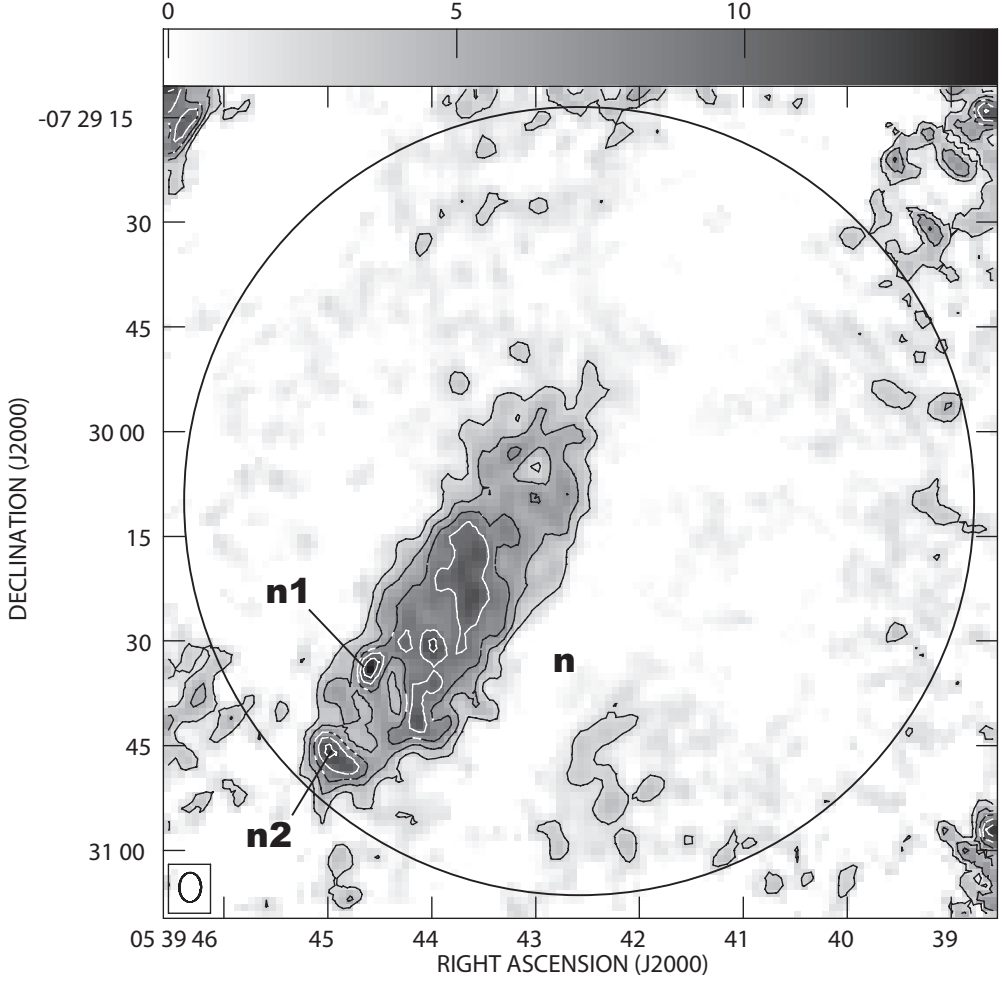}
  \end{center}
  \caption{
NH$_3$ velocity-integrated intensity map of the main NH$_3$ $(J, K)$ = (1, 1) 
component toward TUKH122.
The velocity integration range is from $v_{LSR-K}$ = 3.2 to 4.4 km s$^{-1}$.
The large circle delineates the primary beam of the interferometric observations.
The lowest contour level and contour interval are 2.4 mJy beam$^{-1}$ km s$^{-1}$, which corresponds to 3$\sigma$ and 1.5$\sigma$ at the map center and at the edge of the primary beam, respectively.
The maximum intensity in the primary beam is 14.3 mJy beam$^{-1}$ km s$^{-1}$.
In the lower left corner, the synthesized beam is shown.
}\label{fig:figure4}
\end{figure}

\clearpage

\begin{figure}
\epsscale{.80}
  \begin{center}
    \plotone{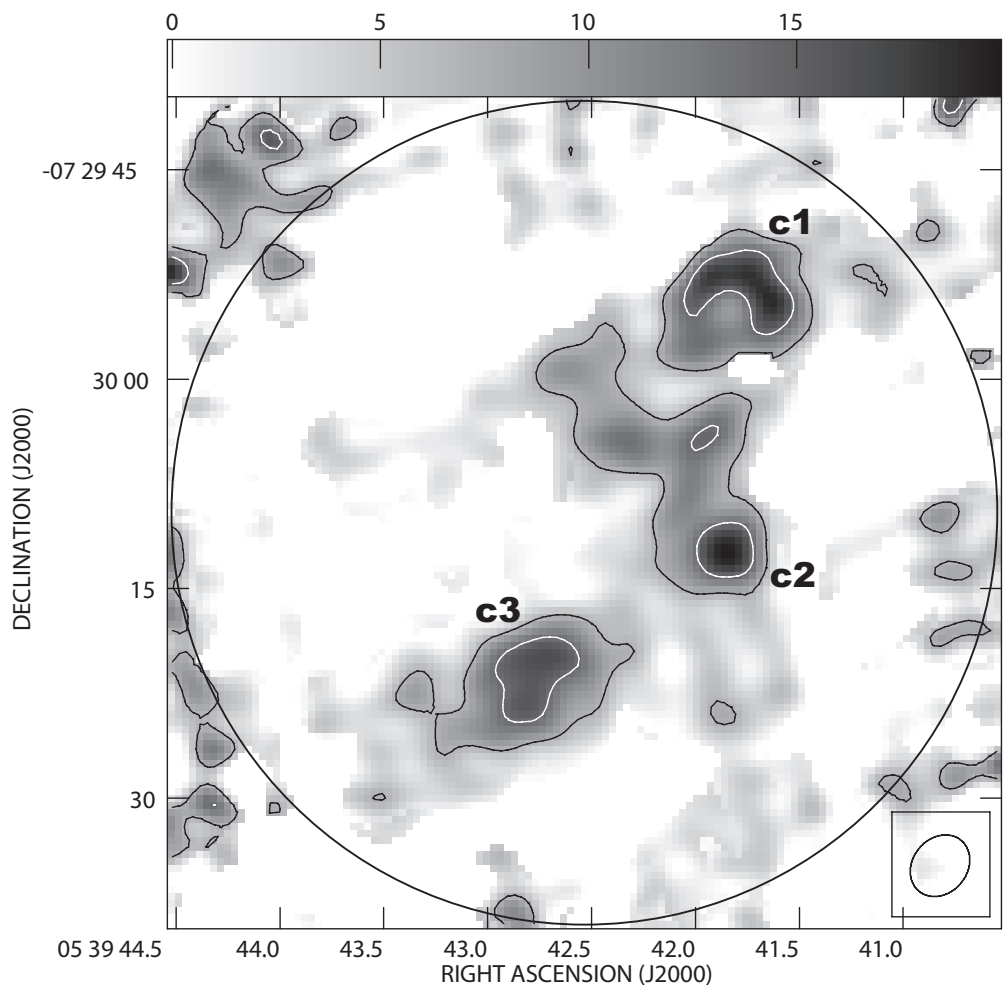}
  \end{center}
  \caption{
CCS velocity-integrated intensity map toward TUKH122.
The velocity integration range is from $v_{LSR-K}$ = 3.6 to 4.0 km s$^{-1}$.
The large circle delineates the primary beam size of the interferometric observations.
The contour interval is 6.9 mJy beam$^{-1}$ km s$^{-1}$, which corresponds to 3$\sigma$ and 1.5$\sigma$ at the map center and at the edge of the primary beam, respectively.
The maximum intensity in the primary beam is 19.8 mJy beam$^{-1}$ km s$^{-1}$.
In the lower right corner, the synthesized beam is shown.
}\label{fig:figure5}
\end{figure}

\clearpage	

\begin{figure}
\epsscale{.80}
  \begin{center}
    \plotone{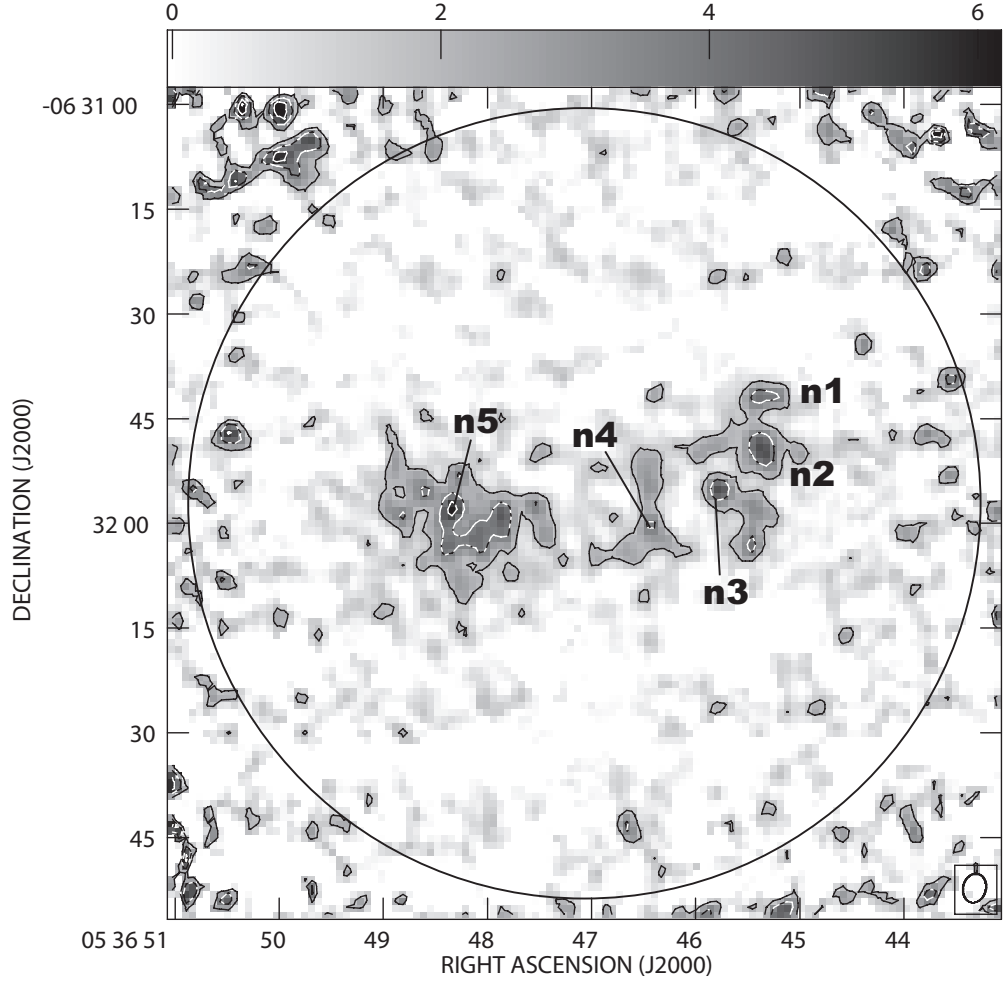}
  \end{center}
  \caption{
NH$_3$ integrated intensity map of the main NH$_3$ $(J, K)$ = (1, 1) 
component toward TUKH083.
The velocity integration range is from $v_{LSR-K}$ = 6.8 to 7.8 km s$^{-1}$.
The contour interval is 1.8 mJy beam$^{-1}$ km s$^{-1}$, which corresponds to 3$\sigma$ and 1.5$\sigma$ at the map center and at the edge of the primary beam, respectively.
The maximum intensity in the prinmary beam is 6.2 mJy beam$^{-1}$ km s$^{-1}$
}\label{fig:figure6}
\end{figure}

\clearpage

\begin{figure}
\epsscale{.80}
  \begin{center}
    \plotone{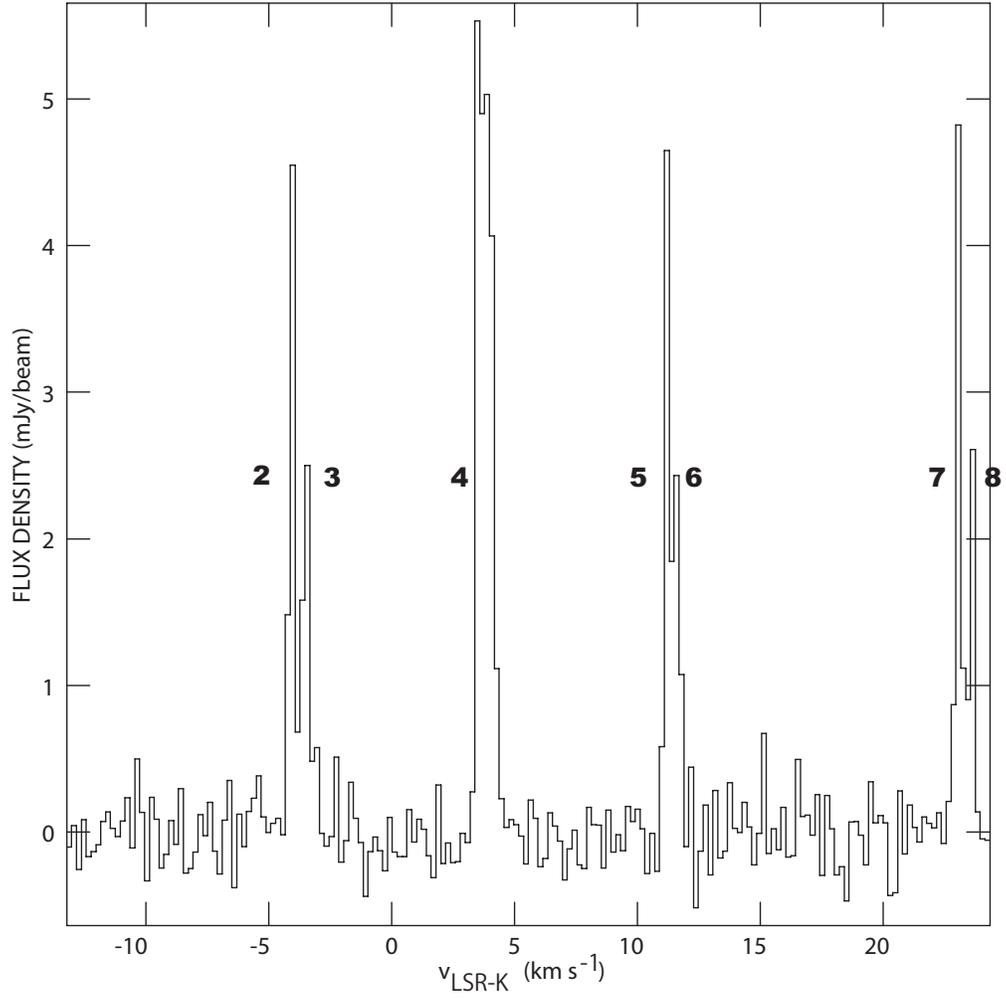}
  \end{center}
  \caption{
NH$_3$ spectrum toward TUKH122.
The abscissa is the LSR-K radial velocity 
with respect to the NH$_3$
($J$,$K$) = (1,1) main component
(feature 4).
The rotated rectangle enclosing TUKH122-n is used for the average spectrum.
}\label{fig:figure7}
\end{figure}

\clearpage
\begin{figure}
\epsscale{.80}
  \begin{center}
    \plotone{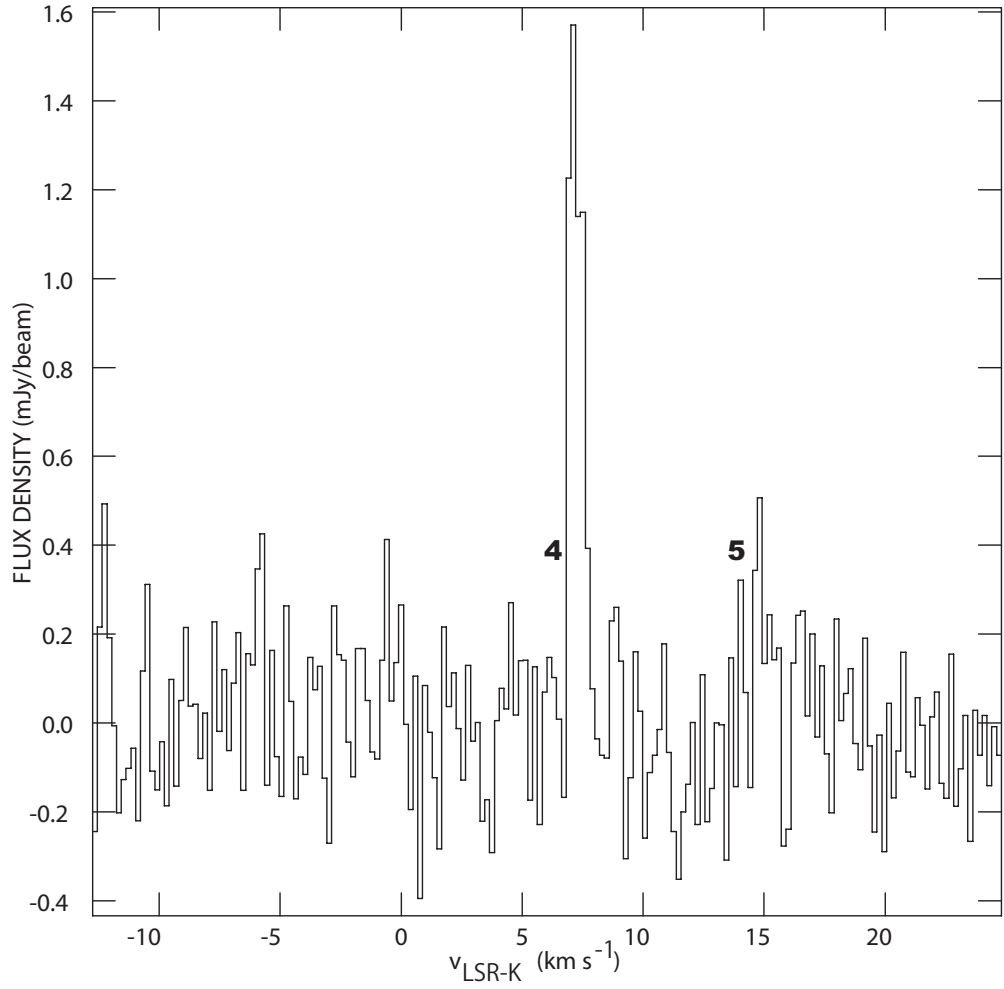}
  \end{center}
  \caption{
The same as Figure 7 but for TUKH083.
The rectangle enclosing TUKH122-n is used for the average spectrum.
NH$_3$ hyperfine feature 1 is located outside the effective spectral window.
}\label{fig:figure8}
\end{figure}

\clearpage

\begin{figure}
\epsscale{.80}
  \begin{center}
    \plotone{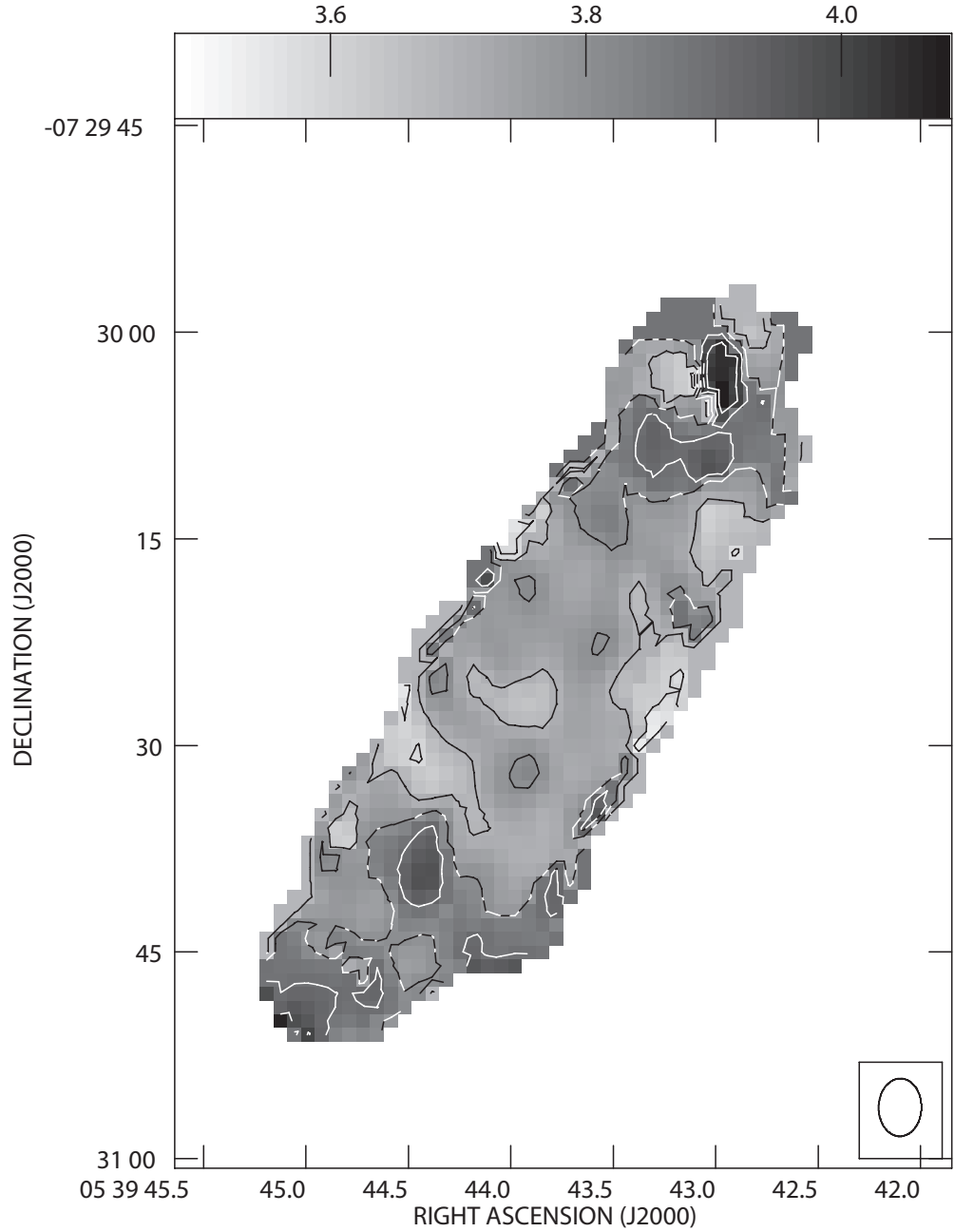}
  \end{center}
  \caption{
NH$_3$ moment 1 (intensity-weighted velocity) map of the NH$_3$ ($J, K$) = (1, 1) feature 4 (main component) toward TUKH122.  
The unit for the velocity is km s$^{-1}$.
Coutours are drawn at  $v_{LSR-K}$ = 3.6, 3.7, 3.8, 3.9, and 4.0 km s$^{-1}$.
The data above 2$\sigma$ at the original channel were used for the moment 1 calculation.
}\label{fig:figure9}
\end{figure}

\clearpage

\begin{figure}
\epsscale{.80}
  \begin{center}
    \plotone{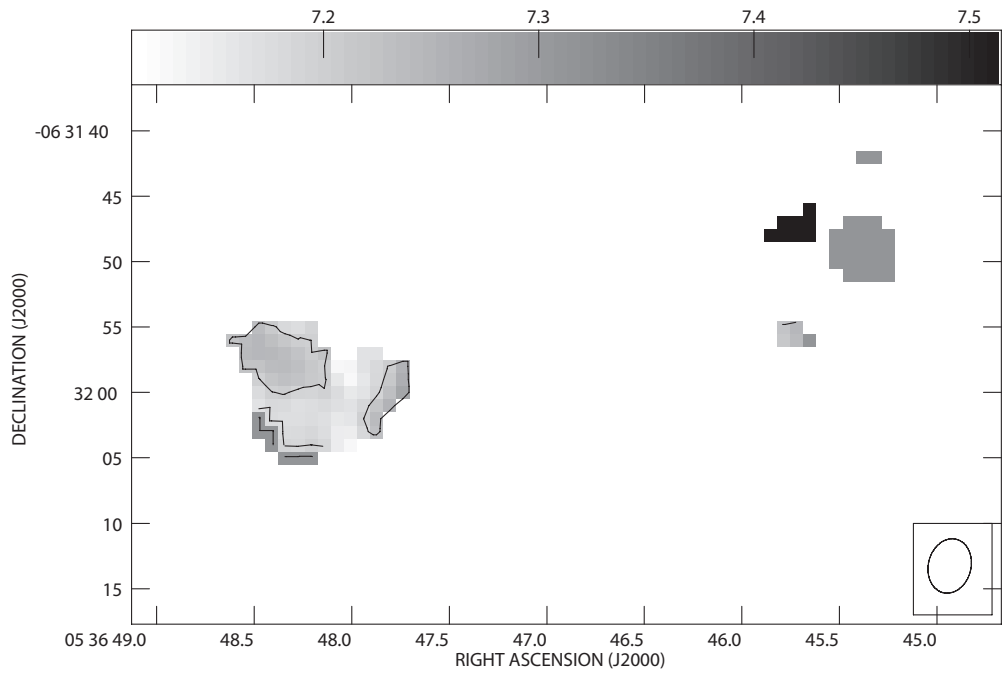}
  \end{center}
  \caption{
The same as Figure 9 but for TUKH083.  
Contours are drawn at $v_{LSR-K}$ = 7.2 and 7.3 km s$^{-1}$.
}\label{fig:figure10}
\end{figure}

\clearpage

\begin{figure}
\epsscale{.80}
  \begin{center}
    \plotone{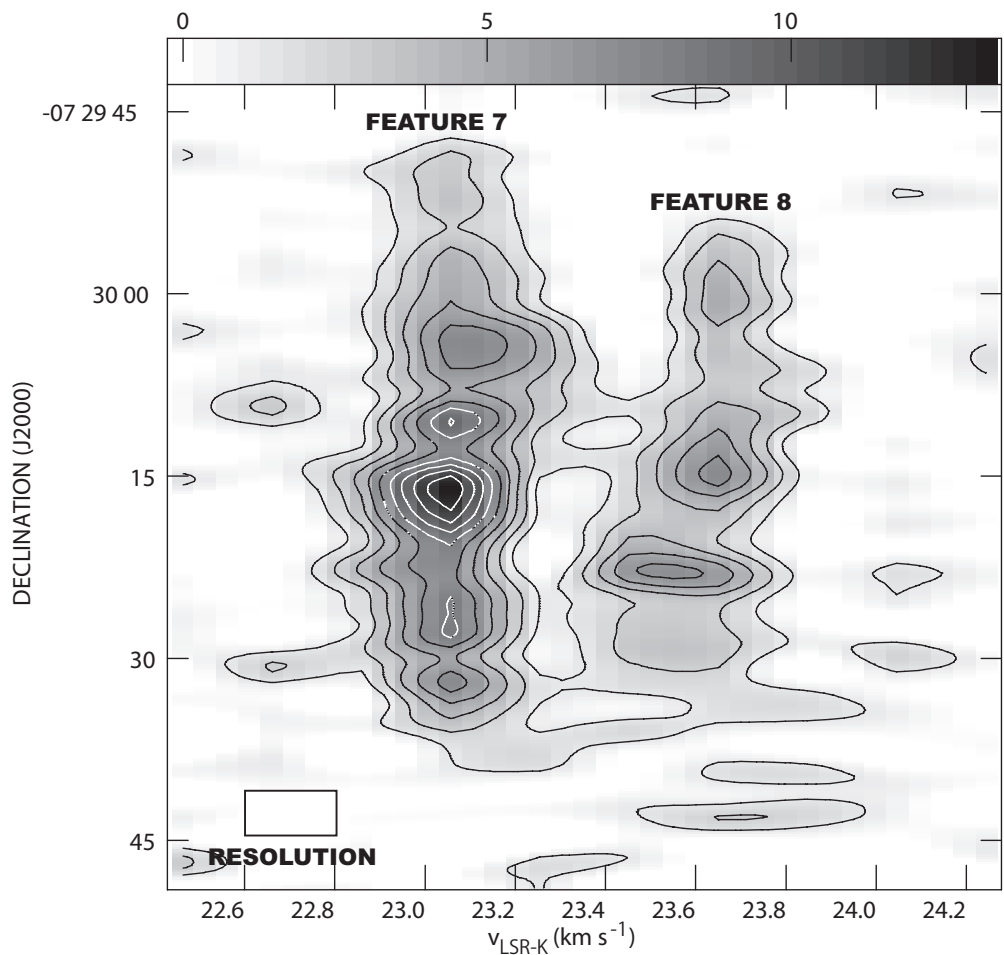}
  \end{center}
  \caption{
The position-velocity diagram for TUKH122 for NH$_3$ hyperfine features 7 and 8.
Contours represent 10, 20, 30, 40, 50, 60, 70, 80, and 90\% of the maximum value
of 13.2 mJy beam$^{-1}$.
The position (declination) is labeled for the eastern side of the 33.3-degree anticlockwisely rotated rectangle enclosing TUKH122-n.
The spectra are averaged across the core elongation.
The radial velocity $v_{LSR-K}$ is labeled for the feature 4.
For features 7 and 8, 19.31 and 19.87 km s$^{-1}$ should be subtracted to obtain the true radial velocity, respectively.}\label{fig:figure11}
\end{figure}

\clearpage

\begin{figure}
\epsscale{.80}
  \begin{center}
    \plotone{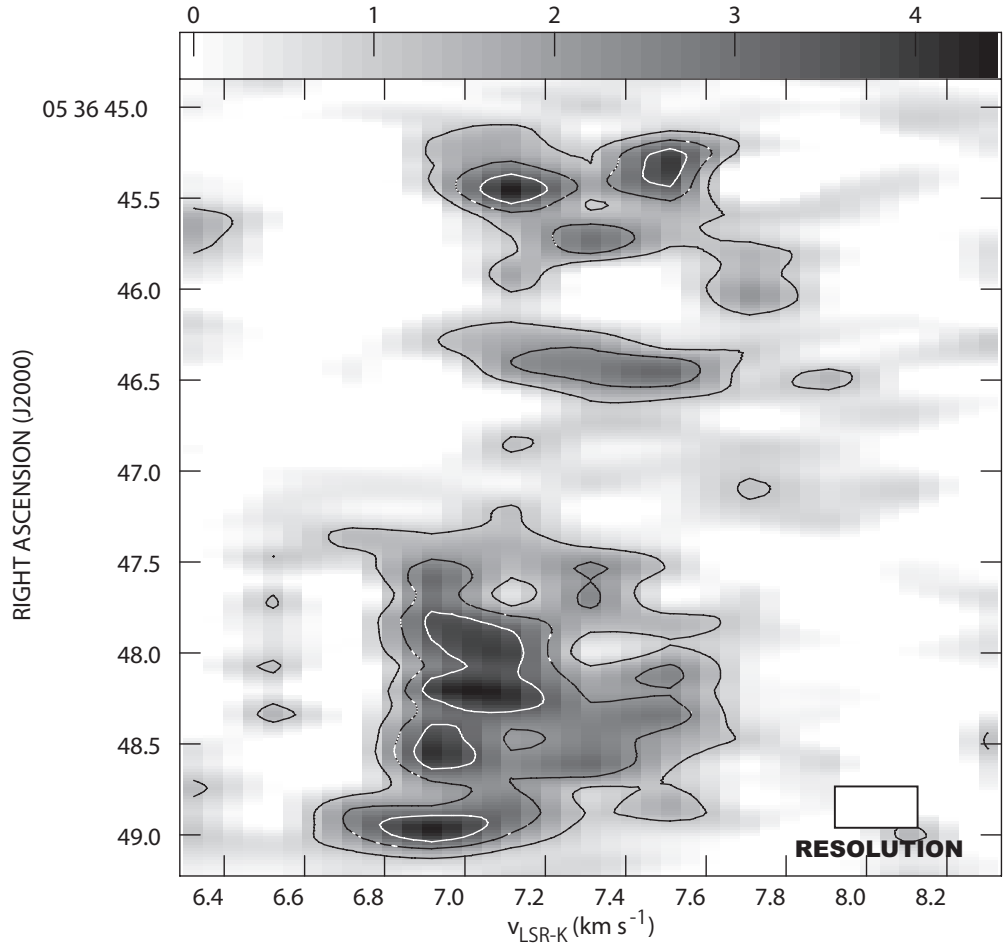}
  \end{center}
  \caption{
NH$_3$ position-velocity diagram along right ascension toward TUKH083 of NH$_3$ hyperfine feature 4.
The spectra are averaged along declination within the rectangle enclosing the
detected NH$_3$ emission.
Contours are drawn at 25\%, 50\%, and 75\%
with respect to the maximum value of 4.4 mJy beam$^{-1}$.
}\label{fig:figure12}
\end{figure}

\clearpage

\begin{figure}
\epsscale{.80}
  \begin{center}
    \plotone{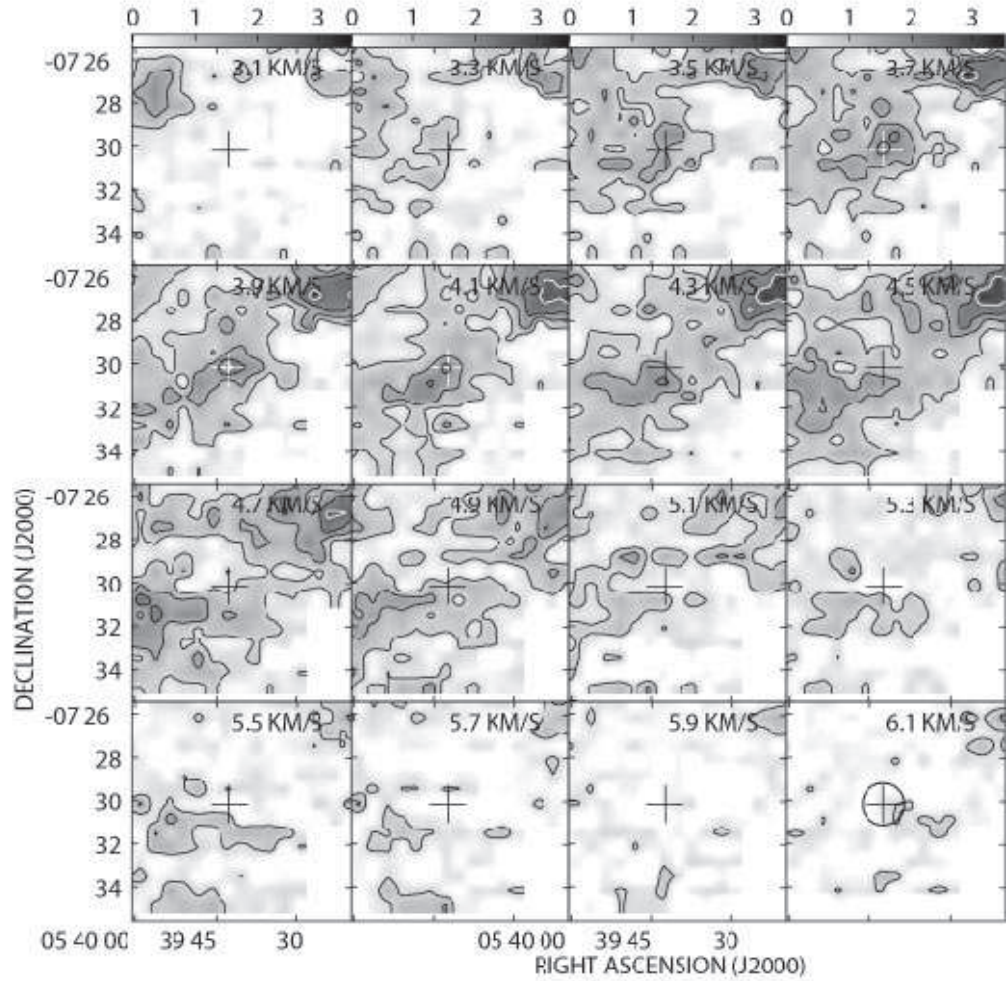}
  \end{center}
  \caption{
CS $J$ = 1$-$0 velocity channel maps of \cite{tat93}.  Contours are drawn from 20\% to 80\% with a step of 20\%
with respect to the maximum value of $T_A^*$ = 3.1 K or $T_R$ = 5.2 K.
The plus sign represents the center of the primary beam size of our NH$_3$ and CCS observations.
The circle at $v_{LSR-K}$ = 6.1 km s$^{-1}$ represents the primary beam size (field of view) for NH$_3$ observations.
}\label{fig:figure13}
\end{figure}

\clearpage






\clearpage

\begin{deluxetable}{llccc}
\tabletypesize{\scriptsize}
\rotate
\tablecaption{Observation Sessions\label{tbl-1}}
\tablewidth{0pt}
\tablehead{
Date  & Observation File Name & Band & Number of Usable Antennas & Used for Image?}
\startdata
2010 August 29   & AT375$\_$sb16969988$\_$1 & Q & 23 & Yes \\
2010 September 2 & AT375$\_$sb1709698    & Q & 25 & Yes \\
2010 September 2 & AT375$\_$sb1709503    & K & 27 & Yes \\
2010 September 7 & AT375$\_$sb1710088$\_$1 & Q & 22 &  Yes \\
2010 September 8 & AT375$\_$sb1709893$\_$1 & Q  & 25 & No \\
2010 September 9 & AT375$\_$sb1697174$\_$1  & K & 24 & No \\
\enddata
\end{deluxetable}

\clearpage

\begin{deluxetable}{ lll }
\tabletypesize{\scriptsize}
\rotate
\tablecaption{ Observational Parameter \label{tbl-1}}
\tablewidth{0pt}
\tablehead{
   &TUKH083                               &TUKH122         \\
}
\startdata
Phase Center RA (J2000.0)  &5$^h$36$^m$47$\fs$0  &5$^h$39$^m$42$\fs$5 \\
Phase Center DEC (J2000.0) &$-$6$\arcdeg$31$\arcmin$56$\arcsec$ &$-$7$\arcdeg$30$\arcmin$09$\arcsec$\\
Synthesized Beam at K       &4$\farcs$2$\times$3$\farcs$3 PA = $-$13$\arcdeg$.8 
                            &4$\farcs$2$\times$3$\farcs$1 PA = $-$1$\arcdeg$.0 \\ 
Synthesized Beam at Q       &4$\farcs$7$\times$3$\farcs$7 PA = $-$33$\arcdeg$.6 
                            &4$\farcs$7$\times$3$\farcs$9 PA = $-$39$\arcdeg$.7\\ 
Primary Beam Size at K & 1$\farcm$90 &1$\farcm$90 \\
Primary Beam Size at Q & 0$\farcm$99 &0$\farcm$99 \\
rms Noise Level at K at 15.625 kHz (0.20 km s${-1}$) resolution &2.3 mJy beam$^{-1}$ &2.6 mJy beam$^{-1}$ \\
rms Noise Level at Q at 31.25 kHz  (0.20 km s${-1}$) resolution &17 mJy beam$^{-1}$ &15 mJy beam$^{-1}$ \\
\enddata
\end{deluxetable}

\clearpage

\begin{deluxetable}{ lcccccccccccc }
\tabletypesize{\scriptsize}
\rotate
\tablecaption{Core and Sub-core\label{tbl-1}}
\tablewidth{0pt}
\tablehead{
Line	&	TUKH	&	Suffix	&		&	RA	&		&		&	DEC	&		&	Flux Density	&	$a_{deconv}$	&	$b_{deconv}$	&	PA$_{deconv}$	\\
	&		&		&	h	&	m	&	s	&	$\arcdeg$	&	$\arcmin$	&	$\arcsec$	&	mJy beam$^{-1}$ km s$^{-1}$	&	$\arcsec$	&	$\arcsec$	&	$\arcdeg$	\\
}
\startdata
NH$_3$	&	083	&	-n1	&	5	&	36	&	45.34 	&	$-$6	&	31	&	42.0 	&	3.5 	&	9.6 	&	4.8 	&	122.5 	\\
NH$_3$	&	083	&	-n2	&	5	&	36	&	45.37 	&	$-$6	&	31	&	49.4 	&	4.5 	&	7.3 	&	5.4 	&	62.2 	\\
NH$_3$	&	083	&	-n3	&	5	&	36	&	45.77 	&	$-$6	&	31	&	55.1 	&	4.9 	&	3.8 	&	1.3 	&	30.4 	\\
NH$_3$	&	083	&	-n4	&	5	&	36	&	46.43 	&	$-$6	&	32	&	0.7 	&	3.5 	&	6.3 	&	2.9 	&	2.9 	\\
NH$_3$	&	083	&	-n5	&	5	&	36	&	48.26 	&	$-$6	&	32	&	0.3 	&	4.2 	&	18.2 	&	13.3 	&	73.7 	\\
NH$_3$	&	122	&	-n	&	5	&	39	&	43.85 	&	$-$7	&	30	&	26.0 	&	10.6 	&	56.3 	&	14.9 	&	147.6 	\\
NH$_3$	&	122	&	-n1	&	5	&	39	&	44.57 	&	$-$7	&	30	&	34.0 	&	13.0 	&	6.3 	&	3.4 	&	163.1 	\\
NH$_3$	&	122	&	-n2	&	5	&	39	&	44.89 	&	$-$7	&	30	&	46.8 	&	11.8 	&	8.5 	&	6.5 	&	50.4 	\\
CCS	&	122	&	-c1	&	5	&	39	&	41.76 	&	$-$7	&	29	&	54.1 	&	18.2 	&	10.8 	&	6.1 	&	118.9 	\\
CCS	&	122	&	-c2	&	5	&	39	&	41.90 	&	$-$7	&	30	&	12.0 	&	18.1 	&	6.3 	&	4.2 	&	60.6 	\\
CCS	&	122	&	-c3	&	5	&	39	&	42.77 	&	$-$7	&	30	&	21.5 	&	17.3 	&	10.0 	&	5.8 	&	126.2 	\\
\enddata
\end{deluxetable}

\clearpage

\begin{deluxetable}{ lccccc }
\tabletypesize{\scriptsize}
\rotate
\tablecaption{ Comparison between Orion cores and L1544\label{tbl-1}}
\tablewidth{0pt}
\tablehead{
	&	TUKH122-n	&	TUKH122 \tablenotemark{a}	&	TUKH123 \tablenotemark{a}	&	L1544 \tablenotemark{b}	&	L1544 \tablenotemark{b}	\\
}
\startdata
$T_k$ (K)	&	$<$ 15 \tablenotemark{c}	&	$<$ 15 \tablenotemark{c}	&	$<$ 15 \tablenotemark{c}	&	9.8 \tablenotemark{d}	&	9.8 \tablenotemark{d}	\\
adopted $T_k$ (K)	&	10	&	10	&	10	&	9.8	&	9.8	\\
Molecule	&	NH$_3$	&	CS $J$ = 1$-$0	&	CS $J$ = 1$-$0	&	N$_2$H$^+$  $J$ = 1$-$0	&	C$^{18}$O $J$ = 1$-$0	\\
$\Delta v$(obs) (km s$^{-1}$)	&	0.20 	&	0.83 	&	1.16 	&	0.37 	&	0.30 	\\
$\Delta v$(tot)  (km s$^{-1}$)	&	0.50 	&	0.94 	&	1.24 	&	0.56 	&	0.52 	\\
$R$ (pc)	&	0.03 	&	0.18 	&	0.23 	&	0.03 	&	0.16 	\\
$M$ ($M_\odot$)	&	1.5 	&	33.0 	&	74.3 	&	2.1 	&	9.2 	\\
$n$ (cm$^{-3}$)	&	2.7E+05	&	2.4E+04	&	2.6E+04	&	2.8E+05	&	9.3E+03	\\
$t_{ff}$ (yr)	&	6.6E+04	&	2.2E+05	&	2.1E+05	&	6.4E+04	&	3.5E+05	\\
$l_J$ (pc)	&	0.03 	&	0.08 	&	0.08 	&	0.02 	&	0.13 	\\
$M_J$ ($M_\odot$)	&	0.3 	&	0.8 	&	0.8 	&	0.2 	&	1.3 	\\
\enddata
\tablenotetext{a}{Tatematsu et al. 1993}
\tablenotetext{b}{Tafalla et al. 1998}
\tablenotetext{c}{Wilson et al. 1999}
\tablenotetext{d}{Benson \& Myers 1989}
\end{deluxetable}




\end{document}